\documentclass[12]{article}
\usepackage{amsmath}
\usepackage{graphicx}
\usepackage{bm}
\begin{document}

\Large
\begin{center}
\bf{A Class of Three-Qubit Contextual Configurations Located in Fano Pentads}
\end{center}
\vspace*{-.3cm}
\begin{center}
Metod Saniga
\end{center}
\vspace*{-.5cm} \normalsize
\begin{center}
Astronomical Institute, Slovak Academy of Sciences\\
SK-05960 Tatransk\' a Lomnica, Slovak Republic\\
(msaniga@astro.sk)  
\end{center}

\vspace*{-.4cm} \noindent \hrulefill

\vspace*{-.1cm} \noindent {\bf Abstract}

\noindent
Given the symplectic polar space of type $W(5,2)$, let us call a set of five Fano planes sharing pairwise a single point a Fano pentad.
Once 63 points of  $W(5,2)$ are appropriately labeled by 63 non-trivial three-qubit observables, any such Fano pentad gives rise
to a quantum contextual set known as Mermin pentagram. Here, it is shown that a Fano pentad also hosts another, closely related contextual set, which features 25 observables and 30 three-element contexts. Out of 25 observables, ten are such that each of them is on six contexts, while each of the remaining 15 observables belongs to two contexts only. Making use of the recent classification of Mermin pentagrams (Saniga et al., Symmetry 12 (2020) 534), it was found that 12,096 such contextual sets comprise 47 distinct types, falling into eight families according to the number ($3, 5, 7, \ldots, 17$) of negative contexts. \\

\vspace*{-.2cm}

%\vspace*{-.2cm}
%\noindent
%{\bf MSC:} 05B25 -- 51E20 

\vspace*{-.2cm} \noindent \hrulefill

\section{Introduction}
Let us call a set of pairwise commuting observables whose product is $+I_d$ or $-I_d$, $I_d$ being the identity, a positive or negative context, respectively. A quantum contextual configuration is a set of contexts such that (i) each observable occurs in an {\it even} number of contexts and (ii) the number of negative contexts is {\it odd}. Any such configuration provides a(n observable-based) proof of the famous Kochen-Specker theorem \cite{ks,spec}. 
There exist a number of proofs of this theorem based on the $N$-qubit Pauli group, $N \geq 2$ (see, e.\,g., \cite{wa12,wa13}). Of them, particularly interesting are those where the structure of underlying symplectic polar space $W(2N-1,2)$  (see, e.\,g., \cite{sp07,hos,thas}) can be invoked to better understand their complex nature.  This idea was recently employed \cite{shj} to get a deeper insight into the structure of the aggregate of 12,096 Mermin pentagrams of the three-qubit symplectic polar space $W(5,2)$. In this note we perform a similar analysis on a closely related class of three-qubit contextual sets having the same number of elements.

\section{Three-Qubit Observables, $\bm{W(5,2)}$ and Fano Pentads}
\begin{figure}[h]
\centerline{\includegraphics[width=13.cm,clip=]{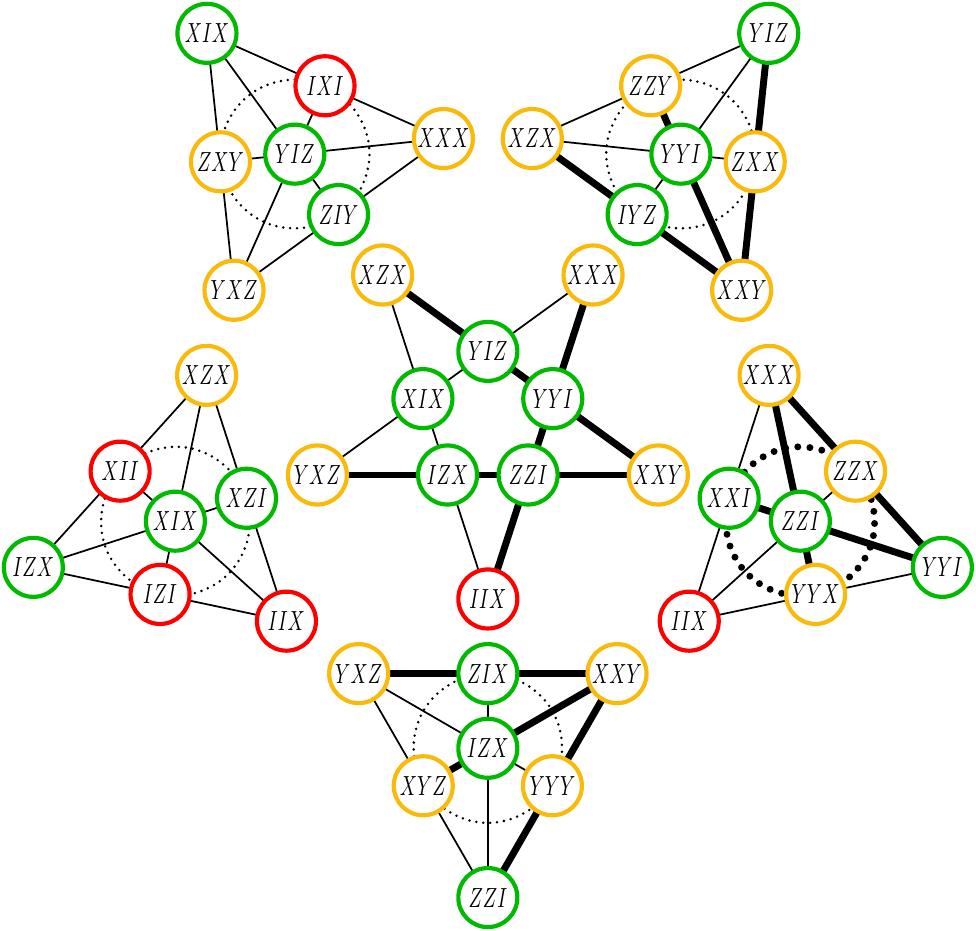}}
\caption{A Fano pentad and the corresponding Mermin pentagram. The Fano plane at the bottom corresponds to the horizontal edge of the pentagram; the remaining correspondences follow readily from the rotational symmetry of the figure. Following Ref.\,8, the three different types of observables are distinguished by different coloring and negative contexts are represented by thick lines; also, $G_1 \otimes G_2 \otimes G_3$ is short-handed 
to $G_1 G_2 G_3$.}
\end{figure}

The three-qubit observables we will be dealing with belong to the set
\begin{eqnarray*}
{\cal S}_3 = \{G_1 \otimes G_2 \otimes G_3:~ G_j \in \{I, X, Y, Z \},~ j \in \{1, 2, 3\}\} \backslash \{I \otimes I \otimes I\}
\end{eqnarray*}
where
\begin{eqnarray*}
I = \left(
\begin{array}{rr}
1 & 0 \\
0 & 1 \\
\end{array}
\right),~
X = \left(
\begin{array}{rr}
0 & 1 \\
1 & 0 \\
\end{array}
\right),~
Y = \left(
\begin{array}{rr}
0 & -i \\
i & 0 \\
\end{array}
\right)
~{\rm and}~
Z = \left(
\begin{array}{rr}
1 & 0 \\
0 & -1 \\
\end{array}
\right).
\end{eqnarray*}
The relevant symplectic polar space $W(5,2)$ can be viewed as the ordinary five-dimensional projective space of order two, PG$(5,2)$, endowed with a non-degenerate symplectic form $\sigma(x,y)$, with its lines and planes being identical to those lines and planes of PG$(5,2)$ on which $\sigma(x,y)$
 vanishes identically; $W(5,2)$ features 63 points, 315 lines and 135 planes, with three points on a line, three planes through a line and both 15 lines and 15 planes through a point. The 63 elements of ${\cal S}_3$ can be put into a one-to-one correspondence with the 63 points of the symplectic polar space $W(5,2)$ in such a way that two commuting elements are collinear  and a maximum set of mutually commuting elements lie in a Fano plane (see, e.\,g., \cite{sp07,hos,thas}). If we take a coordinate basis of $W(5,2)$ in which the symplectic form $\sigma(x,y)$ is given by
\begin{eqnarray*}
\sigma(x,y) =
(x_1 y_4 - x_4 y_1) + (x_2 y_5 - x_5 y_2) + (x_3 y_6 - x_6 y_3),
\end{eqnarray*}
then this correspondence has the form
\begin{eqnarray*}
G_j \leftrightarrow (x_j, x_{j+3}),~j \in \{1, 2, 3\},
\end{eqnarray*}
assuming that
\begin{eqnarray*}
I \leftrightarrow (0,0),~X \leftrightarrow (0,1),~Y \leftrightarrow (1,1),~Z \leftrightarrow (1,0);
\end{eqnarray*}
thus, for example, the point of $W(5,2)$ having coordinates $(0,1,1,1,1,0)$ corresponds to the element $X \otimes Y \otimes Z$. An important structural property of $W(5,2)$ is a set of five Fano planes such that their pairwise intersections are all different and consist of a single point each, the shared points forming in each Fano plane an affine plane of order two. We shall call such a set of Fano planes a {\it Fano pentad} and the affine planes consisting of shared points will be referred to as distinguished ones. 
    
\section{A Notable Class of MP-Related Contextual Sets}  
Let us consider now the `three-qubit' $W(5,2)$, i.\,e. $W(5,2)$ having its points labeled by three-qubit observables as described in the preceding section, and call its line/plane positive or negative according as the product of observables located on it is $+I \otimes I \otimes I$ or $-I \otimes I \otimes I$, respectively. As first noticed in \cite{sale}, and further elaborated in \cite{shj}, any Fano pentad in such space gives rise to a unique Mermin pentagram whose contexts (edges) correspond to the five distinguished affine planes. This property is also illustrated in Figure 1; the points of the pentagram and the Fano planes are labeled by three-qubit observables in such a way that the distinguished affine planes are obtained by the removal of those lines (and points located on them) that are represented by dotted circles. It is interesting to see that, given a Fano pentad, we can also obtain a contextual configuration if we remove from each Fano plane this distinguished line, but {\it keep} the points located on it and regard each of the remaining six lines as a context. Such a configuration  consists of 25 observables and 30 contexts, where each of those ten observables that are also on the pentagram belongs to six contexts and each of the remaining 15 observables is shared by two contexts only. Using the language of Waegell and Aravind \cite{wa13}, our configuration carries the symbol $10_6 15_2 - 30_3$. From this construction it is obvious that not only is there a unique $10_6 15_2 - 30_3$ configuration for each Mermin pentagram, but the two kinds of configurations are so closely related to each other that  we can readily classify our $10_6 15_2 - 30_3$ configurations making use of the strategy and results of Ref.\,8. In our classification, each type will be characterized by the same string of parameters as used for Mermin pentagrams \cite{shj}, i.\,e. the number of negative contexts, distribution of types of observables and partitioning of types of valu(at)ed Fano planes, adding one more essential characteristic -- the type of the Mermin pentagram accommodated in the same Fano pentad. 

To this end, let us recall \cite{shj} that a three-qubit observable from ${\cal S}_3$ is of type $A$, $B$ or $C$  depending on whether it features two $I$'s, one $I$ or no $I$, respectively. Next, a Fano plane of the three-qubit $W(5,2)$ can be positive or negative. As all Fano planes contain just three observables
of type $B$ lying on the same line, they can only differ from each other in the corresponding affine part. A negative Fano plane has all four affine observables of type $C$ and contains three concurrent negative lines. The affine part of a positive Fano plane consists either of a single observable of type $A$ and three observables of type $C$, or {\it vice versa}. In the former case one distinguishes between the case whether the plane contains negative lines (type $a$) or not (type $b$), the latter case being type $c$; if a positive Fano plane contains negative lines, there are always four, no three of them being concurrent. Our analysis of $10_6 15_2 - 30_3$ configurations was performed in the following steps: we picked up a representative
Mermin pentagram of a given type (see Table 1 of Ref.\,8), found the corresponding Fano pentad, localized in this pentad our configuration and read-off
its parameters. The principal results of our analysis are collected in Table 1. Figure 1 can also serve as an example of this procedure for a Mermin pentagram of type 15. The associated Fano pentad comprises two negative Fano planes (top-right and bottom ones) and three positive ones; the middle-right plane is of type $a$, the top-left one is of type $b$ and the middle-left of type $c$. Removing from this pentad the five lines represented by dotted circles yields the corresponding $10_6 15_2 - 30_3$ configuration.
\begin{table}[pth!]
\begin{center}
\caption{A `group-geometric' classification of $10_6 15_2 - 30_3$ configurations. Following the nomenclature of Ref.\,8, column one ($T$) shows the type, column two ($C^{-}$) the number of negative contexts in a configuration of the given type, columns three to five ($O_{A}$ to $O_{C}$) indicate the number of observables of individual types, column six ($F^{-}$) the number of negative Fano planes and columns seven to eight ($F^{+}_{a}$ to $F^{+}_{c}$) the distribution of types of positive Fano planes in the Fano pentad. The last column ($T^{\cal{P}}$) indicates the type of associated Mermin pentagrams.  } \vspace*{0.9cm}
%\resizebox{\columnwidth}{!}{% 
\scalebox{0.8}{
\begin{tabular}{|r|c|ccc|c|ccc|r|}
\hline \hline
$T$        & $C^{-}$       & $O_{A}$ & $O_{B}$ & $O_{C}$  & $F^{-}$        & $F^{+}_{a}$       & $F^{+}_{b}$       & $F^{+}_{c}$  & $T^{\cal{P}}$   \\
\hline
  1        & 17            & 2       & 11      & 12       & 3              & 2                 & 0                 & 0    & 5 \\
	\hline
  2        & 15            & 0       & 15      & 10       & 5              & 0                 & 0                 & 0    & 1 \\
  3        & 15            & 1       & 15      &  9       & 3              & 2                 & 0                 & 0    & 2 \\
	\hline
  4        & 13            & 0       & 11      & 14       & 5              & 0                 & 0                 & 0    & 4 \\
  5        & 13            & 1       & 10      & 14       & 4              & 1                 & 0                 & 0    & 21 \\
	6        & 13            & 1       & 11      & 13       & 3              & 2                 & 0                 & 0    & 9 \\
	7        & 13            & 2       & 11      & 12       & 3              & 1                 & 1                 & 0    & 6 \\
	8        & 13            & 3       & 10      & 12       & 2              & 2                 & 1                 & 0    & 22 \\
	\hline
  9        & 11            & 1       & 10      & 14       & 4              & 0                 & 1                 & 0    & 3 \\
  10       & 11            & 2       & 10      & 13       & 2              & 2                 & 1                 & 0    & 14 \\
	11       & 11            & 2       & 11      & 12       & 3              & 1                 & 1                 & 0    & 24 \\
	12       & 11            & 3       & 11      & 11       & 3              & 1                 & 0                 & 1    & 10 \\
	13       & 11            & 4       & 10      & 11       & 2              & 2                 & 0                 & 1    & 30 \\
	14       & 11            & 5       & 11      &  9       & 1              & 2                 & 1                 & 1    & $28b$ \\
	\hline
	15       & 9             & 1       & 11      &  13      & 3              & 0                 & 2                 & 0    & 11 \\
	16       & 9             & 2       & 10      &  13      & 2              & 1                 & 2                 & 0    & 31 \\
	17       & 9             & 2       & 11      &  12      & 3              & 0                 & 2                 & 0    & 7 \\
	18       & 9             & 2       & 11      &  12      & 1              & 2                 & 2                 & 0    & 17 \\
	19       & 9             & 3       & 10      &  12      & 2              & 1                 & 2                 & 0    & 23 \\
	20       & 9             & 3       & 11      &  11      & 3              & 0                 & 1                 & 1    & 12 \\
	21       & 9             & 4       & 10      &  11      & 2              & 1                 & 1                 & 1    & 15 \\
	22       & 9             & 4       & 10      &  11      & 2              & 1                 & 1                 & 1    & 32 \\
	23       & 9             & 4       & 11      &  10      & 1              & 2                 & 1                 & 1    & 18 \\
	24       & 9             & 4       & 11      &  10      & 1              & 2                 & 1                 & 1    & 36 \\
	25       & 9             & 5       & 10      &  10      & 2              & 1                 & 0                 & 2    & 16 \\
  26       & 9             & 1       & 15      &  9       & 3              & 0                 & 2                 & 0    & 8 \\
  27       & 9             & 5       & 11      &  9       & 3              & 0                 & 0                 & 2    & 13 \\
  28       & 9             & 5       & 11      &  9       & 1              & 2                 & 0                 & 2    & 20 \\
  29       & 9             & 5       & 11      &  9       & 1              & 2                 & 1                 & 1    & $28a$ \\
  30       & 9             & 3       & 15      &  7       & 1              & 2                 & 1                 & 1    & 19 \\
\hline  
 31        & 7             & 1       & 11      &  13      & 3              & 0                 & 2                 & 0    & 25 \\
 32        & 7             & 3       & 11      &  11      & 3              & 0                 & 1                 & 1    & 26 \\
 33        & 7             & 4       & 11      &  10      & 1              & 1                 & 2                 & 1    & $37b$ \\
 34        & 7             & 5       & 10      &  10      & 2              & 1                 & 0                 & 2    & 34 \\
 35        & 7             & 5       & 11      &  9       & 3              & 0                 & 0                 & 2    & 27 \\
 36        & 7             & 6       & 10      &  9       & 0              & 2                 & 1                 & 2    & 41 \\
\hline
 37        & 5             & 4       & 10      &  11      & 2              & 0                 & 2                 & 1    & 33 \\
 38        & 5             & 4       & 11      &  10      & 1              & 1                 & 2                 & 1    & $37a$ \\
 39        & 5             & 5       & 10      &  10      & 2              & 0                 & 1                 & 2    & 35 \\
 40        & 5             & 5       & 11      &  9       & 1              & 1                 & 1                 & 2    & 39 \\
 41        & 5             & 6       & 11      &  8       & 1              & 1                 & 0                 & 3    & 43 \\
\hline
 42        & 3             & 5       & 11      &  9       & 1              & 0                 & 3                 & 1    & 29 \\ 
 43        & 3             & 5       & 11      &  9       & 1              & 0                 & 2                 & 2    & 40 \\
 44        & 3             & 6       & 10      &  9       & 0              & 1                 & 2                 & 2    & 42 \\
 45        & 3             & 6       & 11      &  8       & 1              & 0                 & 1                 & 3    & 44 \\
 46        & 3             & 3       & 15      &  7       & 1              & 0                 & 3                 & 1    & 38 \\
 47        & 3             & 6       & 15      &  4       & 1              & 0                 & 0                 & 4    & 45 \\
 \hline \hline
\end{tabular}
}
%}%
\end{center}
\end{table}
From the figure we read off that our configuration has nine negative contexts and features four observables of type $A$, ten observables of type $B$ and 11 of type $C$; hence, it belongs to type 21.  

Although there is one-to-one correspondence between the set of Mermin pentagrams and that of our $10_6 15_2 - 30_3$ configurations, the numbers of types are different, 45 versus 47, respectively. The origin of this discrepancy is rather simple. There are two particular types of Mermin pentagrams, namely types 28 and 37 (see Table 1 of \cite{shj}), whose internal structure is `felt' by the corresponding configurations. A Mermin pentagram of type 28 has a single observable of type $A$ and a single negative context, and the associated $10_6 15_2 - 30_3$ configuration is sensitive  on whether the observable of type $A$
does ($28a$) or does not ($28b$) lie in the negative context. Similarly, a Mermin pentagram of type 37 has also only one negative context but two observables of type $A$, and the  associated $10_6 15_2 - 30_3$ configuration discriminates whether this negative context does ($37a$) or does not ($37b$) contain one of these two observables. On the other hand, we can also find a couple of examples where the opposite holds, i.\,e. where configurations `originating' from pentagrams of different types have identical parameters;  one example is furnished by types 21 and 22, the other entails types 23 and 24. There are also worth mentioning some other features of our taxonomy readily discernible from Table 1. If a configuration has just a single observable of type $A$, then the positive Fano planes from its pentad are of only one type. If a configuration features two observables of type $A$, then it does not exhibit a positive Fano plane of type $c$. And a configuration endowed with four observables of type $A$ has just a single positive Fano plane of type $c$. One further observes that a configuration with an even (odd) number of observables of type $B$ is characterized by an even (odd) number of negative Fano planes. It is also an interesting feature that there is no $10_6 15_2 - 30_3$ configuration with a single negative context, or having more that 17 ones.

\section{Conclusion}
We have discovered a new remarkable class of 12,096 three-qubit quantum contextual configurations of type $10_6 15_2 - 30_3$ that live, like well-known Mermin pentagrams, in Fano pentads. Due to their close relation with Mermin pentagrams, their classification followed rather straightforwardly from the taxonomy of pentagrams and, in addition, led to a discovery of finer structure of two particular types of pentagrams. As in the case of Mermin pentagrams, also here a key element of our analysis was making use of the structure of the associated symplectic polar space $W(5,2)$. 
Our final remark concerns the occurrence of the number 12,096, which may serve as a revival of an older hypothesis \cite{psh} about the possible role of the split Cayley hexagon of order two, a distinguished subgeometry of $W(5,2)$ whose automorphism group has the same number of elements, in three-qubit quantum contextuality issues.

\section*{Acknowledgments}
This work was supported by the  Slovak VEGA Grant Agency, Project $\#$ 2/0004/20. We also thank Zsolt Szab\'o (Budapest University of Technology and Economics) for the help with the preparation of the figure.


\vspace*{-.1cm}
\begin{thebibliography}{10}
\itemsep=-2pt
\bibitem{ks}
S. Kochen and E. P. Specker, The problem of hidden variables in quantum mechanics, Journal of Mathematics and Mechanics 17 (1967) 59. 
\bibitem{spec}
E. P. Specker, Die Logik nicht gleichzeitig entscheidbarer Aussagen, Dialectica 14 (1960)  239.
\bibitem{wa12}
M. Waegell and P. K. Aravind, Proofs of the Kochen-Specker theorem based on a system of three qubits, Journal of Physics A: Mathematical and 
Theoretical 45 (2012) Art. No. 405301.
\bibitem{wa13}
M. Waegell and P. K. Aravind, Proofs of the Kochen-Specker theorem based on the $N$-qubit Pauli group, Physical Review A 88 (2013) Art. No. 012102.
\bibitem{sp07}
M. Saniga and M. Planat, Multiple qubits as symplectic polar spaces of order two, Advanced Studies in Theoretical Physics 1 (2007) 1.
\bibitem{hos}
H. Havlicek, B. Odehnal and M. Saniga, Factor-group-generated polar spaces and (multi-)qudits, Symmetry, Integrability and Geometry: Methods and Applications 5 (2009) Art. No. 096. 
\bibitem{thas}
K. Thas, The geometry of generalized Pauli operators of $N$-qudit Hilbert space, EPL -- Europhysics Letters 86 (2009) Art. No. 60005. 
\bibitem{shj}
M. Saniga, F. Holweck and H. Jaffali, Taxonomy of three-qubit Mermin pentagrams, Symmetry 12 (2020) Art. No. 534.
\bibitem{sale}
M. Saniga and P. L\'evay, Mermin's pentagram as an ovoid of PG(3,2), EPL -- Europhysics Letters 97 (2012) Art. No. 50006. 
\bibitem{psh}
M. Planat, M. Saniga and F. Holweck, Distinguished three-qubit `magicity' via automorphisms of the split Cayley hexagon, Quantum Information Processing 12 (2013) 2535. 
\end{thebibliography}
\end{document}